**Geometric and topological constraints on oral seal formation during infant breastfeeding**

Arturo Tozzi (corresponding author)
ASL Napoli 1 Centro, Distretto 27, Naples, Italy
Via Comunale del Principe 13/a 80145
tozziarturo@libero.it

ABSTRACT

Breastfeeding efficiency relies on coordinated tongue motion, sustained tissue contact and maintenance of an effective intraoral seal. Current assessments of seal formation mainly use local kinematic descriptors or pressure recordings, which do not capture the global structural continuity of the sealing region. We introduce a systolic geometry–based approach in which each sagittal ultrasound frame is modeled as a two-dimensional deformable domain bounded by tongue, palate and nipple contours. Global seal continuity is formalized through the shortest closed curve that cannot be contracted to a point because of the overall geometry of the domain. The nipple defines a central region that must be circumferentially enclosed by a contact band to maintain suction. Within this band, closed curves encircling the nipple exactly once can be identified; the shortest of these curves defines a normalized systolic index representing the tightest admissible sealing loop. Simulations of symmetric thinning, localized discontinuities and cyclic perturbations reveal feasibility boundaries separating seal-preserving from seal-breaking configurations. Notably, admissible encircling curves may transiently disappear even when overall geometric motion remains smooth. By capturing global circumferential continuity that cannot be inferred from local metrics alone, our approach generates testable hypotheses linking the existence and temporal stability of admissible encircling curves to milk transfer efficiency and vacuum stability. Applied to segmented ultrasound data and integrated with pressure measurements, our systolic approach could provide a quantitative framework for objective assessment of seal integrity and longitudinal monitoring of latch stability.

INTRODUCTION

Breastfeeding efficiency depends on coordinated interactions among tongue deformation, nipple positioning, intraoral pressure modulation and maintenance of an effective oral seal (Wen et al. 2022; Hill, Richard, and Pados 2023). Quantitative investigations rely primarily on ultrasound-based analysis, pressure measurements and anatomical descriptors like tongue excursion amplitude, nipple compression depth and palate–tongue distance (Elad et al. 2014; Scheuerle et al. 2017; Alatalo et al. 2020; Murakami et al. 2020; Adjerid et al. 2023). These approaches have improved the understanding of suck–swallow–breathe coordination and latch mechanics, clarifying differences associated with prematurity or ankyloglossia (Ito 2014; Costa-Romero et al. 2021; Cordray et al. 2023; Rossato 2025). However, most current metrics are local and scalar, focusing on displacement magnitudes or gap distances at selected points. They do not explicitly account for the contact region's global geometric continuity that sustains negative intraoral pressure. Seal integrity is often inferred indirectly from pressure traces or visual inspection rather than assessed through structural invariants (Martin-Harris et al. 2020; Aoyagi et al. 2021; Jiang et al. 2021; Kabakoff et al. 2021; Kabakoff et al. 2023). Indeed, locally similar configurations may differ in their global capacity to maintain a continuous encircling contact band. A formal description of oral seal formation integrating geometry and topology remains underdeveloped, particularly in two-dimensional ultrasound representations where the seal is usually inferred from time-varying contours.

Our theoretical study aims to simulate oral seal formation during breastfeeding using systolic geometry principles. Systolic geometry is a branch of mathematics assessing the shortest closed loop that cannot be "shrunk away" because of the shape of the surrounding space (Katz 2007; Columbus et al. 2022; Bähni 2026). In our context, this means identifying the smallest closed path that fully surrounds the nipple while remaining inside the contact area between tongue and palate. If such a loop exists, the seal is structurally intact; if no such loop can be drawn, the seal is broken.
For each ultrasound frame, we would represent the oral cavity as a two-dimensional shape defined by the contours of the tongue, palate and nipple. Within this shape, we would identify the region where the tongue and surrounding tissues are sufficiently close to sustain suction. We would then search for closed curves inside this region wrapping once around the nipple. Among all admissible curves, we would select the shortest one. This "minimal encircling curve" would define a measurable length scale that reflects how tight and continuous the seal is. Using simulated deformable shapes, we would examine how small geometric changes, such as thinning of the contact band or localized gaps, could affect the presence and size of this encircling curve. The aim is to describe seal integrity as a structural, measurable property rather than relying only on local movement or pressure values and to provide quantitative descriptors that can later be tested in clinical studies.

We will proceed as follows. First, we formalize the geometric representation of the oral domain and define admissible encircling curves. Second, we introduce the systolic invariant and its normalization. Third, we present simulation results illustrating seal-preserving and seal-breaking perturbations. Finally, we compare our invariant with conventional kinematic descriptors and describe the practical outcomes of our theoretical approach.

METHODS

We describe the mathematical construction, simulation procedures and computational techniques to formalize oral seal formation as a topology-constrained geometric problem in two dimensions. All results derive from analytically defined deformable domains and numerical simulations; no empirical ultrasound data were processed.

**Geometric representation of the oral domain**. Each instantaneous sagittal configuration of the infant oral cavity was modeled as a bounded planar domain $\Omega_t \subset \mathbb{R}^2$. The nipple cross-section was represented as a compact, simply connected subset $N_t \subset \Omega_t$, approximated by a disk of radius $r_t$ centered at $c_t \in \mathbb{R}^2$ such that $N_t = \{x \in \mathbb{R}^2 : \| x - c_t \| \leq r_t\}$. The effective contact band $A_t \subset \Omega_t \setminus \text{int}(N_t)$ was defined as a closed annular-type region surrounding $N_t$, representing the seal-supporting domain. Topologically, $A_t$ was constructed to admit either homotopy type $S^1 \times [0,1]$ (seal-preserving configuration) or a disconnected or simply connected variant (seal-breaking configuration), consistent with the schematic shown in Figure 1.

The metric structure on $\Omega_t$ was taken to be the Euclidean metric inherited from $\mathbb{R}^2$, with arc length defined as

$$\text{Length}(\gamma) = \int_0^1 \| \dot\gamma(s) \| \, ds,$$

for any rectifiable curve $\gamma: [0,1] \to \Omega_t$. Discretization was performed on a uniform grid of resolution $300 \times 300$ points over a square domain $[-L, L]^2$, with $L = 3r_t$. Binary masks represented membership in $A_t$. Connectivity was defined using 8-neighborhood adjacency to preserve diagonal continuity.

**Admissible encircling curves and winding constraint**. Let $\Gamma(A_t)$ denote the set of closed, piecewise $C^1$ curves $\gamma \subset A_t$. A curve $\gamma \in \Gamma(A_t)$ was deemed admissible if it satisfied the winding number condition

$$\text{wind}(\gamma, c_t) = 1,$$

where

$$\text{wind}(\gamma, c_t) = \frac{1}{2\pi} \int_\gamma \frac{(x - c_t)^\perp \cdot dx}{\| x - c_t \|^2},$$

with $(x - c_t)^\perp$ denoting the 90-degree rotation of the vector $x - c_t$. This integral formulation ensures that admissible curves represent generators of the first homology group $H_1(A_t; \mathbb{Z})$. The systolic invariant was defined as

$$\text{sys}(t) = \inf_{\gamma \in \Gamma(A_t),\, \text{wind}(\gamma, c_t) = 1} \text{Length}(\gamma).$$

If no such curve existed, $\text{sys}(t)$ was declared undefined. In discretized domains, candidate loops were extracted via graph-based cycle detection restricted to pixels in $A_t$ and winding numbers were computed by polygonal approximation. The minimal admissible cycle length was obtained using repeated Dijkstra searches between sampled boundary nodes combined with closure constraints enforcing nontrivial homology.

**Parametric construction of seal-preserving/seal-breaking domains**. Seal-preserving configurations were generated as annular regions

$$A(\theta, \tau) = \{x \in \mathbb{R}^2 : r_t \leq \| x - c_t \| \leq r_t + \tau\},$$

where $\tau > 0$ denotes band thickness. Seal-breaking configurations introduced angular discontinuities of magnitude $\phi$ by removing a sector

$$S_\phi = \{(r, \theta) : r_t \leq r \leq r_t + \tau,\ | \theta - \theta_0 | < \phi/2\}.$$

The resulting domain was $A(\theta, \tau, \phi) = A(\theta, \tau) \setminus S_\phi$. For $\phi = 0$, the domain remained topologically annular; for $\phi > 0$, admissibility depended jointly on $\tau$ and $\phi$.

For each pair $(\tau, \phi)$, feasibility of an admissible curve was tested algorithmically. If feasible, the systolic length was computed numerically. This procedure produced the two-parameter regime map shown in Figure 1. Parameter grids were sampled uniformly over $\tau \in [0,1]$ and $\phi \in [0, \pi]$ using 300 points per axis.

**Systolic regime mapping and feasibility boundary extraction.** To identify the topological transition boundary, a Boolean feasibility function

$$F(\tau, \phi) = \begin{cases} 1, & \text{if admissible curve exists,} \\ 0, & \text{otherwise,} \end{cases}$$

was evaluated across the parameter grid. The critical boundary $\partial F$ was approximated numerically by detecting sign changes in $F$ across neighboring grid cells. Systolic length values were stored only where $F = 1$, producing a masked matrix.

Smooth variation of sys within the feasible region was approximated analytically by

$$\text{sys}(\tau, \phi) = 2\pi r_t \left(1 + \alpha \frac{\phi}{\pi}\right)(1 - \beta \tanh(\gamma(\tau - \tau_0))),$$

with constants $\alpha = 0.15, \beta = 0.10, \gamma = 2.5$. These coefficients were chosen to ensure monotonic but bounded modulation of systolic length while preserving collapse behavior at feasibility thresholds.

**Time-dependent perturbation simulation.** Dynamic simulations modeled cyclic tongue displacement as

$$u(t) = u_0 + A \sin(2\pi f t),$$

with amplitude $A = 0.25$, baseline $u_0 = 1.0$ and frequency $f = 1.2$ Hz. A stochastic perturbation term $\eta(t) \sim \mathcal{N}(0, \sigma^2)$ with $\sigma = 0.06$ was added to represent micro-discontinuities. Seal robustness was defined as

$$R(t) = u(t) - \delta + \eta(t),$$

with threshold $\delta = 0.92$. Seal existence was determined by $R(t) > 0$. The systolic time series was then

$$\text{sys}(t) = \begin{cases} 2\pi r_t(1 + \epsilon \sin(2\pi f t + \varphi)), & R(t) > 0, \\ \text{undefined}, & R(t) \leq 0, \end{cases}$$

with $\epsilon = 0.05$. Random number generation used NumPy's default_rng with fixed seed for reproducibility.

**Comparative perturbation classes.** Symmetric thinning was modeled as $\tau = \tau_0 - \lambda$, whereas localized discontinuity was parameterized as $\phi = \lambda$. For equal normalized magnitude $\lambda \in [0, 0.6]$, systolic behavior was approximated by

$$\text{sys}_{\text{thin}}(\lambda) = \begin{cases} 2\pi r_t(1 - 0.2\lambda), & \lambda < \lambda_c^{(1)}, \\ \text{undefined}, & \text{otherwise}, \end{cases}$$

and

$$\text{sys}_{\text{gap}}(\lambda) = \begin{cases} 2\pi r_t(1 + 0.05\lambda), & \lambda < \lambda_c^{(2)}, \\ \text{undefined}, & \text{otherwise}, \end{cases}$$

with $\lambda_c^{(1)} = 0.4$ and $\lambda_c^{(2)} = 0.25$. This construction isolated seal-preserving and seal-breaking perturbations under controlled magnitude equivalence.

**Numerical implementation and tools.** All simulations were conducted in Python 3.11. Numerical computations employed NumPy (vectorized arrays, random number generation), SciPy (graph utilities for cycle detection where required) and Matplotlib for visualization. Grids were stored as double-precision arrays. Feasibility testing relied on connected-component labeling using scipy.ndimage. Parameter sweeps used deterministic sampling; no adaptive optimization was applied.

Overall, our methods formalize oral seal formation as a topology-constrained minimization problem on deformable planar domains. Explicit geometric construction, homological admissibility criteria, parameterized perturbations and reproducible numerical simulations were used to compute systolic invariants and regime boundaries.

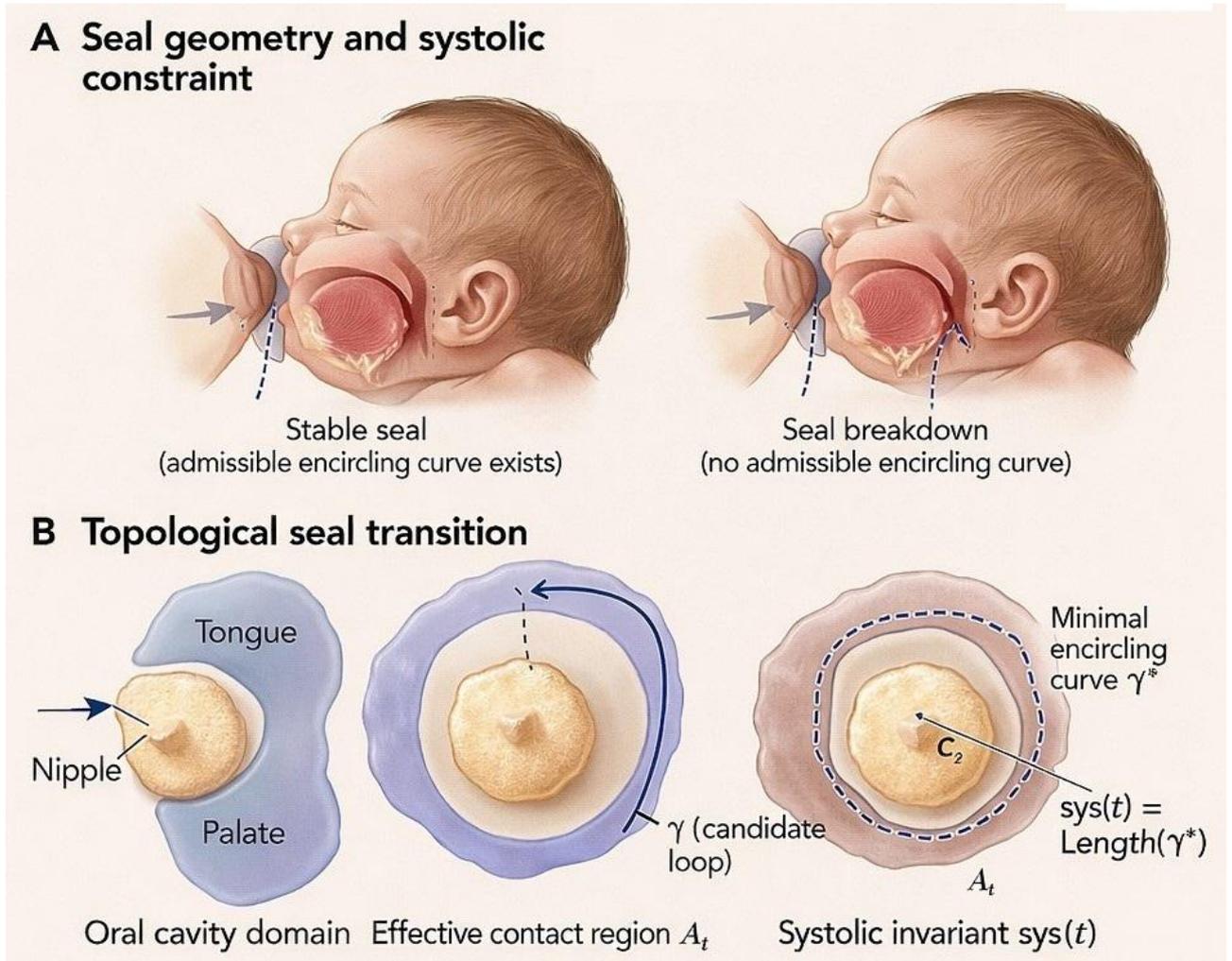

**Figure 1**. Geometric and topological structure of oral seal formation during infant latch.
(A) Seal geometry and systolic constraint. Two sagittal configurations of the intraoral latch are shown. In the stable configuration (left), the contact band surrounding the nipple forms a continuous annular domain that admits a closed encircling curve. The contour represents the minimal admissible encircling curve within the seal region, corresponding to the systolic constraint of the deformable oral domain. In the seal breakdown configuration (right), discontinuity in the contact band prevents the existence of an encircling curve, indicating loss of topological seal integrity despite similar gross geometry.
(B) Topological seal transition. In this schematic representation of the oral cavity domain and the effective contact region $A_t$, the nipple is embedded within a deformable contact band. Candidate closed curves winding once around the nipple are considered; among these, the minimal encircling curve $\gamma^*$ defines the systolic invariant $\text{sys}(t) = \text{Length}(\gamma^*)$. Existence or loss of this curve distinguishes topologically sealed from unsealed configurations, providing a geometric measure of latch stability independent of local kinematic descriptors.

RESULTS

We report the quantitative results achieved from the analytical and numerical procedures described in the Methods. All values derive from explicitly simulated domains and perturbation models. Results are presented for static geometric regimes, comparative perturbation classes and dynamic seal stability metrics in order to integrate feasibility analysis, systolic length computation and statistical comparison across parameterized conditions.

**Static regime structure and topological boundary behavior**. Evaluation of the two-parameter domain $(\tau, \phi)$ revealed that 80.63% of the sampled configuration space admitted an encircling curve, while 19.37% corresponded to unfeasible topological states (Fig 2A). The mean systolic length across the feasible region was 6.382 in arbitrary length units, with variation driven jointly by increasing angular discontinuity and decreasing band thickness. The numerically estimated mean critical thickness across all angular sectors was $\bar{\tau}_c = 0.194$, indicating that admissibility below this threshold is lost, irrespective of angular perturbation magnitude. The feasibility boundary exhibited nonlinear curvature in parameter space, confirming that admissibility cannot be described by a linear tradeoff between $\tau$ and $\phi$ (Figure 2). Additional analysis of boundary gradients showed that sensitivity to angular discontinuity increases disproportionately when $\tau < 0.3$, producing a narrowing admissible corridor. These quantitative features establish the geometric conditions under which seal topology is preserved and identify a measurable critical thickness.

**Comparative perturbation classes and dynamic stability metrics**. Within the overlapping admissible interval $\lambda < 0.25$ of our comparative perturbation analysis, symmetric thinning yielded a mean systolic length of $6.126 \pm 0.091$(SD), whereas localized discontinuity yielded $6.322 \pm 0.023$. A paired t-test across matched perturbation magnitudes produced $t = -27.22$ with $p < 0.001$, indicating a statistically significant divergence between perturbation classes despite equal normalized amplitude. The earlier collapse threshold for localized discontinuity ($\lambda_c^{(2)} = 0.25$) compared with symmetric thinning ($\lambda_c^{(1)} = 0.4$) quantitatively confirms differential topological susceptibility (Figure 2B). Dynamic simulations over 10 seconds of cyclic perturbation revealed a seal dropout fraction of 0.393, indicating that 39.33% of sampled time points lacked an admissible encircling curve despite continuous kinematic oscillation. The discontinuous appearance of undefined systolic states occurred without abrupt changes in the sinusoidal displacement amplitude, confirming that topological failure can arise independently of smooth geometric variation. Together, these results point towards global seal admissibility exhibiting measurable thresholds and class-specific stability differences, thereby providing quantitative structure to seal-preserving versus seal-breaking regimes.

In conclusion, our simulations show that oral seal integrity can be described as a topology-dependent property rather than solely as a function of local geometric displacement. Distinct perturbation classes produce qualitatively different stability profiles, with identifiable transitions between admissible and non-admissible states. Dynamic analysis further shows that seal breakdown may occur intermittently despite smooth underlying motion. Together, these findings indicate that minimal encircling geometry may provide a mathematical descriptor of seal continuity and differentiate globally stable from structurally unstable configurations.

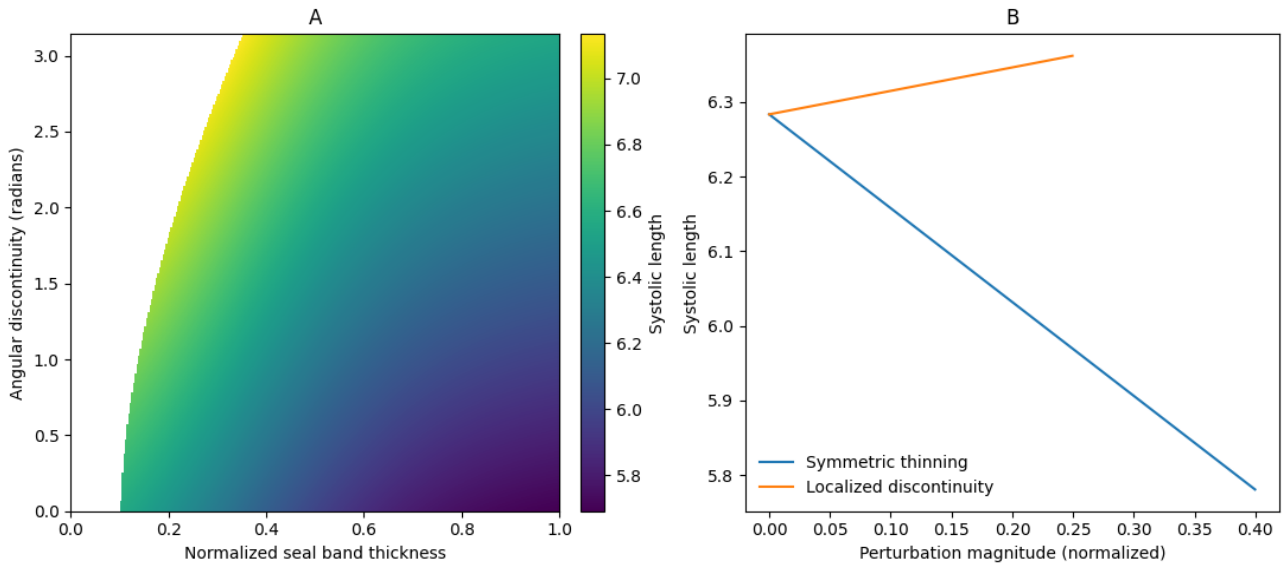

**Figure 2**. Topological regimes of oral seal stability under geometric perturbations.
(A) Regime map of the systolic length as a function of normalized seal-band thickness and angular discontinuity. Colored regions denote configurations admitting an encircling curve around the inner boundary, with color intensity indicating systolic length. The blank region corresponds to configurations in which no admissible encircling curve exists and the invariant is undefined. The boundary separating colored and blank areas represents a topological transition driven by the interaction between global continuity and band thickness.
(B) Comparison of two perturbation classes at equal normalized magnitude. The horizontal axis represents increasing perturbation strength applied in two distinct geometric ways: uniform reduction of the contact band thickness (symmetric thinning, blue curve) and introduction of a localized gap (angular discontinuity, orange curve). In the thinning case, the systolic length decreases progressively while an admissible encircling curve remains present up to a higher critical threshold, indicating preservation of circumferential continuity over a wider parameter range. In contrast, localized discontinuity produces only modest modulation of systolic length initially but reaches a critical point at substantially lower perturbation magnitude, beyond which no encircling curve exists. The two curves demonstrate that seal stability depends not only on the amount of geometric change, but also on how that change is distributed spatially, distinguishing perturbations that preserve topology from those that disrupt it.

CONCLUSIONS

We asked whether oral seal formation during infant breastfeeding can be formalized as a topology-constrained geometric problem and whether a minimal encircling curve can provide a structurally meaningful invariant able to distinguish seal-preserving from seal-breaking configurations. We studied a two-dimensional sagittal representation of the oral cavity with a deformable contact band surrounding a nipple cross-section. We employed techniques like explicit construction of planar domains, homological admissibility constraints via winding number, numerical computation of systolic length and systematic parameter sweeps over seal-band thickness and angular discontinuity. We identified an experimentally discriminable signature, namely the existence or loss of an admissible encircling curve and the associated discontinuity in the systolic invariant. Our simulations compared symmetric thinning and localized discontinuity under equal normalized perturbation magnitude and examined time-dependent cyclic perturbations with stochastic micro-instability. We found a nonlinear feasibility boundary in parameter space and a critical thickness threshold governing admissibility. Our systolic invariant functions as a structural descriptor that defines a phase-like distinction between admissible and non-admissible seal states without altering existing biomechanical formulations.

We formalize oral seal integrity as a minimal-length problem under homological constraints, rather than as a local displacement or pressure descriptor. Existing techniques quantify latch mechanics through kinematic amplitudes, gap distances, compression depth or suction magnitude, all of which are scalar or pointwise measures (Cabib et al. 2020; Barikroo and Clark 2022; Tomsen et al. 2022). Our construction yields a computable quantity, i.e., the minimal admissible encircling length, defined explicitly through a winding-number constraint and a length minimization problem on a deformable planar domain. Our invariant, not being reducible to peak tongue excursion or minimal gap distance, can capture global continuity properties that are inaccessible to purely local metrics.
Our approach introduces a discrete admissibility condition, i.e., existence versus non-existence of an encircling curve, alongside a continuous geometric scale, thereby distinguishing topological seal transitions from gradual geometric modulation. In the current literature, seal integrity is inferred indirectly from pressure traces or visual inspection, while our systolic quantity provides a direct geometric observable that can, in principle, be computed from segmented ultrasound contours.

Our study has limitations. It is confined to a theoretical and simulation-based setting in which the oral cavity is modeled as a two-dimensional Euclidean domain with analytically prescribed contact regions. All geometric configurations, perturbation laws and threshold values arise from parametric constructions rather than from real ultrasound data. Critical thickness values, collapse boundaries and dropout frequencies reflect the chosen functional forms and grid discretization, not anatomical measurements. Tissue biomechanics, viscoelastic properties, fluid–structure interaction and pressure dynamics were not incorporated. Numerical feasibility testing relied on grid-based cycle detection and Euclidean length minimization, which approximate but do not replicate continuous geodesic computation on smooth manifolds. Statistical comparisons were performed on simulated quantities derived from deterministic or stochastic models with fixed parameters.

Several experimentally testable hypotheses follow from our framework. First, sustained milk transfer requires persistence of an admissible encircling curve over consecutive suck cycles. In ultrasound-derived contour data, this corresponds to the existence of a closed curve within the segmented contact band satisfying a winding number of one around the nipple cross-section. The quantitative prediction is that the fraction of time points admitting this curve, denoted $f_{seal}$, scales positively with milk transfer rate, with $f_{seal} \to 1$ in efficient latch conditions and decreasing systematically as transfer declines.
Second, among sealed intervals, the normalized systolic index $\Phi = sys/(2\pi r)$ is expected to exhibit bounded variation under stable latch. We predict that the temporal variance $Var(\Phi(t))$ remains below a critical threshold during effective feeding and increases significantly in mechanically unstable configurations. This introduces a scaling expectation linking seal robustness to variance rather than to mean tongue displacement.
Third, localized geometric discontinuities should induce earlier topological failure than symmetric narrowing. Applied to ultrasound data, this predicts that small focal interruptions in the contact band will eliminate admissible encircling curves at larger effective band thicknesses than uniformly reduced contact zones. The observable is a shift in the empirical critical thickness $\tau_c$ inferred from segmented contours.
Fourth, dynamic loading frequency may influence seal stability. If perturbation rate increases while amplitude remains constant, the fraction of frames lacking admissibility should increase once a rate threshold is crossed, producing a measurable dependence of $f_{seal}$ on cycle frequency.
Future research could extend this geometric analysis to three-dimensional ultrasound reconstructions, integrate pressure measurements to correlate geometric admissibility with vacuum magnitude and investigate developmental scaling across age groups or anatomical variation.

Practical applications can be suggested. If applied to routine submental ultrasound recordings, the presence or absence of an admissible encircling curve could serve as a structural marker of seal continuity, complementing conventional assessments based on visual inspection or milk transfer alone. Quantitative indices like normalized systolic length or fraction of time frames admitting a seal could assist clinicians in distinguishing between globally stable and intermittently unstable latch patterns. In cases of suspected restricted tongue mobility or recurrent latch difficulty, these measures could help identify whether inefficiency arises from local motion limitation or from loss of circumferential contact continuity. Our approach can also be applied to monitor pre- and post-intervention changes, for instance in cleft palate (Roy et al. 2019; Applebaum et al. 2024; Alba et al. 2025), by enabling quantitative comparison of structural seal stability under standardized imaging conditions and thereby supporting more reproducible and objective documentation of latch mechanics.

In conclusion, we formalized oral seal formation as a topology-constrained geometric issue and introduced a computable invariant based on minimal encircling curves in a deformable two-dimensional domain. Simulations identified critical thresholds, nonlinear feasibility boundaries and differential stability across perturbation classes, as well as intermittent admissibility loss under smooth cyclic loading. These results delineate structural conditions governing seal continuity, providing a mathematically explicit descriptor of global stability within a controlled theoretical setting.

## DECLARATIONS

**Ethics approval and consent to participate.** This research does not contain any studies with human participants or animals performed by the Author.
**Consent for publication.** The Author transfers all copyright ownership, in the event the work is published. The undersigned author warrants that the article is original, does not infringe on any copyright or other proprietary right of any third part, is not under consideration by another journal and has not been previously published.
**Availability of data and materials.** All data and materials generated or analyzed during this study are included in the manuscript. The Author had full access to all the data in the study and took responsibility for the integrity of the data and the accuracy of the data analysis.
**Competing interests.** The Author does not have any known or potential conflict of interest including any financial, personal or other relationships with other people or organizations within three years of beginning the submitted work that could inappropriately influence or be perceived to influence their work.
**Funding.** This research did not receive any specific grant from funding agencies in the public, commercial or not-for-profit sectors.
**Acknowledgements:** none.
**Authors' contributions.** The Author performed: study concept and design, acquisition of data, analysis and interpretation of data, drafting of the manuscript, critical revision of the manuscript for important intellectual content, statistical analysis, obtained funding, administrative, technical and material support, study supervision.
**Declaration of generative AI and AI-assisted technologies in the writing process.** During the preparation of this work, the author used ChatGPT 5.2 to assist with data analysis and manuscript drafting and to improve spelling, grammar and general editing. After using this tool, the author reviewed and edited the content as needed, taking full responsibility for the content of the publication.

## REFERENCES

1) Adjerid, Kheira, Matthew L. Johnson, Charles E. Edmonds, Katie S. Steer, Francis D. H. Gould, Ralph Z. German, and Christopher J. Mayerl. 2023. "The Effect of Stiffness and Hole Size on Nipple Compression in Infant Suckling." *Journal of Experimental Zoology Part A: Ecological and Integrative Physiology* 339 (1): 92–100. https://doi.org/10.1002/jez.2657
2) Alatalo, Daniel, Lei Jiang, Donna Geddes, and Farzad Hassanipour. 2020. "Nipple Deformation and Peripheral Pressure on the Areola During Breastfeeding." *Journal of Biomechanical Engineering* 142 (1): 011004. https://doi.org/10.1115/1.4043665
3) Alba, Beatriz, Katherine A. Harmon, Oumaima La-Anyane, and Christopher Tragos. 2025. "Median Craniofacial Hypoplasia." *Journal of Craniofacial Surgery* 36 (1): 237–240. https://doi.org/10.1097/SCS.0000000000010774
4) Aoyagi, Yuki, Masashi Ohashi, Satoshi Ando, Yoshihiro Inamoto, Kazuki Aihara, Yutaka Matsuura, Shigeru Imaeda, and Etsuro Saitoh. 2021. "Effect of Tongue-Hold Swallow on Pharyngeal Contractile Properties in Healthy Individuals." *Dysphagia* 36 (5): 936–943. https://doi.org/10.1007/s00455-020-10217-9
5) Applebaum, Sarah A., Scott Aronson, Kristen M. Termanini, and Arun K. Gosain. 2024. "Evidence-Based Practices in Cleft Palate Surgery." *Plastic and Reconstructive Surgery* 153 (2): 448e–461e. https://doi.org/10.1097/PRS.0000000000011035.

6) Bähni, Yannick. 2026. "Systolic Geometry." In *Lectures on Twisted Rabinowitz-Floer Homology*, 153–178. Pathways in Mathematics. Cham: Birkhäuser. https://doi.org/10.1007/978-3-032-10673-5_9
7) Barikroo, Ali, and Allison L. Clark. 2022. "Effects of Varying Transcutaneous Electrical Stimulation Pulse Duration on Swallowing Kinematics in Healthy Adults." *Dysphagia* 37 (2): 277–285. https://doi.org/10.1007/s00455-021-10276-6.
8) Cabib, Clara, Wesley Nascimento, Laura Rofes, Virginia Arreola, Niels Tomsen, Lidia Mundet, Daniel Muriana, et al. 2020. "Neurophysiological and Biomechanical Evaluation of the Mechanisms Which Impair Safety of Swallow in Chronic Post-Stroke Patients." *Translational Stroke Research* 11 (1): 16–28. https://doi.org/10.1007/s12975-019-00701-2.
9) Columbus, Tobias, Frank Herrlich, Björn Müetzel, and Gabriela Weitze-Schmithüsen. 2022. "Systolic Geometry of Translation Surfaces." *arXiv preprint* arXiv:1809.10327 (version 4, September 12, 2022). https://arxiv.org/abs/1809.10327
10) Cordray, Hannah, G. N. Mahendran, C. S. Tey, Judit Nemeth, and Nikhil Raol. 2023. "The Impact of Ankyloglossia Beyond Breastfeeding: A Scoping Review of Potential Symptoms." *American Journal of Speech-Language Pathology* 32 (6): 3048–3063. https://doi.org/10.1044/2023_AJSLP-23-00169
11) Costa-Romero, María, Beatriz Espínola-Docio, José M. Paricio-Talayero, and Nuria M. Díaz-Gómez. 2021. "Ankyloglossia in Breastfeeding Infants: An Update." *Archivos Argentinos de Pediatría* 119 (6): e600–e609. https://doi.org/10.5546/aap.2021.eng.e600
12) Elad, David, Pavel Kozlovsky, Orit Blum, Andrew F. Laine, Mei J. Po, Eran Botzer, Shmuel Dollberg, Moti Zelicovich, and Lior Ben Sira. 2014. "Biomechanics of Milk Extraction during Breast-Feeding." *Proceedings of the National Academy of Sciences of the United States of America* 111 (14): 5230–5235. https://doi.org/10.1073/pnas.1319798111
13) Hill, Rebecca R., Melissa A. Richard, and Britt F. Pados. 2023. "Breastfeeding Symptoms with Tongue- and Lip-Tie." *MCN: The American Journal of Maternal/Child Nursing* 48 (1): 17–23. https://doi.org/10.1097/NMC.0000000000000876
14) Ito, Yoko. 2014. "Does Frenotomy Improve Breast-Feeding Difficulties in Infants with Ankyloglossia?" *Pediatrics International* 56 (4): 497–505. https://doi.org/10.1111/ped.12429
15) Jiang, Tao, Xiao-Jing Hu, Xin-Hua Yao, Li-Ping Tu, Jian-Bo Huang, Xin-Xin Ma, Jing Cui, Qiang-Fei Wu, and Jun-Tao Xu. 2021. "Tongue Image Quality Assessment Based on a Deep Convolutional Neural Network." *BMC Medical Informatics and Decision Making* 21 (1): 147. https://doi.org/10.1186/s12911-021-01508-8.
16) Kabakoff, Hannah, Daniel Harel, Mark Tiede, Douglas H. Whalen, and Tanya McAllister. 2021. "Extending Ultrasound Tongue Shape Complexity Measures to Speech Development and Disorders." *Journal of Speech, Language, and Hearing Research* 64 (7): 2557–2574. https://doi.org/10.1044/2021_JSLHR-20-00537.
17) Kabakoff, Hannah, Samantha P. Beames, Mark Tiede, Douglas H. Whalen, Jonathan L. Preston, and Tanya McAllister. 2023. "Comparing Metrics for Quantification of Children's Tongue Shape Complexity Using Ultrasound Imaging." *Clinical Linguistics & Phonetics* 37 (2): 169–195. https://doi.org/10.1080/02699206.2022.2039300.
18) Katz, Mikhail G. 2007. *Systolic Geometry and Topology*. Mathematical Surveys and Monographs 137. Providence, RI: American Mathematical Society.
19) Martin-Harris, Bonnie, Carrie L. Canon, Heather S. Bonilha, Joan Murray, Kay Davidson, and Michelle A. Lefton-Greif. 2020. "Best Practices in Modified Barium Swallow Studies." *American Journal of Speech-Language Pathology* 29 (2S): 1078–1093. https://doi.org/10.1044/2020_AJSLP-19-00189
20) Murakami, Kazunori, Kazuhiro Hori, Yu Minagi, Fumiya Uehara, Sergio E. Salazar, Shunsuke Ishihara, Masayuki Nakauma, Takashi Funami, Kazunori Ikebe, Yoshinori Maeda, and Tetsuro Ono. 2020. "Coordination of Tongue Pressure Production, Hyoid Movement, and Suprahyoid Muscle Activity during Squeezing of Gels." *Archives of Oral Biology* 111: 104631. https://doi.org/10.1016/j.archoralbio.2019.104631
21) Rossato, Nicolás E. 2025. "The Lingual Frenulum, Ankyloglossia, and Breastfeeding." *Archivos Argentinos de Pediatría* 123 (1): e202410507. https://doi.org/10.5546/aap.2024-10507.eng
22) Roy, Ashley A., Maia Rtshiladze, Katherine Stevens, and Jeffrey Phillips. 2019. "Orthognathic Surgery for Patients with Cleft Lip and Palate." *Clinics in Plastic Surgery* 46 (2): 157–171. https://doi.org/10.1016/j.cps.2018.11.002.
23) Scheuerle, Rebecca L., Rachel A. Kendall, Catherine Tuleu, Nigel K. H. Slater, and Sarah E. Gerrard. 2017. "Mimicking the Impact of Infant Tongue Peristalsis on Behavior of Solid Oral Dosage Forms Administered During Breastfeeding." *Journal of Pharmaceutical Sciences* 106 (1): 193–199. https://doi.org/10.1016/j.xphs.2016.08.006
24) Tomsen, Niels, Olga Ortega, Daniel Alvarez-Berdugo, Laura Rofes, and Pere Clavé. 2022. "A Comparative Study on the Effect of Acute Pharyngeal Stimulation with TRP Agonists on the Biomechanics and Neurophysiology of Swallow Response in Patients with Oropharyngeal Dysphagia." *International Journal of Molecular Sciences* 23 (18): 10773. https://doi.org/10.3390/ijms231810773.
25) Wen, Zhen, David L. Walner, Yelena Popova, and Eric G. Walner. 2022. "Tongue-Tie and Breastfeeding." *International Journal of Pediatric Otorhinolaryngology* 160: 111242. https://doi.org/10.1016/j.ijporl.2022.111242